# A New Normalization Form for Limited Distinct Attributes

Niko S. Snell

Department of Engineering Science and Biomedical Engineering, University of Auckland, nikosnell17148@gmail.com

Rayen C. Lee

Department of Electrical, Computer, and Software Engineering, University of Auckland

In modern databases, the practice of data normalization continues to be important in improving data integrity, minimizing redundancies, and eliminating anomalies. However, since its inception and consequent improvements, there have been no attempts to document a method which constrains the values of attributes capable of only possessing a limited quantity of values. These non-limited distinct attributes pose a problem throughout many relational databases as they have the potential to cause data anomalies and query inaccuracies. **Thus,** a new database normalization method, Limited Distinct Normal Form (LDNF), is necessary in order to improve upon the currently established data normalization process. In brief, LDNF is a method which turns non-limited distinct attributes into limited distinct attributes by forcing the attributes to conform to a limited quantity of values. Utilizing LDNF in tandem with existing normal forms fulfills a need in normalization that is otherwise not present when only using current methods. A formal approach to LDNF is therefore proposed.

# 1 INTRODUCTION

In our world today there persists a heavy reliance on data as it continues to be a key influence on all sectors of any developed economy. This notion is admittedly repetitious to the point of being overstated, but remains important to acknowledge at the risk of being cliché. Databases inevitably play a key role in handling this data, storing this critical information for later usage. One such type of database is the relational database, which maintains data in terms of their relationship with one another. Depending on the needs of a given relational database, developers may employ various methods in order to meet end user requirements, the most prominent of which being normalization. Normalization attempts to promote data integrity, minimize data redundancies, and eliminate update, insertion, and deletion anomalies. This is achieved through a series of guidelines classified under normal forms (NFs).

It is apparent that there remains room for further improvement when it comes to normalization and its various NFs as more could be done in order to better secure a database, but also to ensure that data is kept at a satisfactory quality. To this day, there exist no NFs that attempt to minimize errors by enforcing constraints upon the values distinct attributes can hold, and it stands to reason that one should be introduced. For the purpose of this discussion, distinct attributes refers to attributes which can only contain a small, finite set of values, for instance, weather. Whether or not a set of finite values can be considered small is inevitably subjective, but the upper limit of a given set's size will ultimately depend on the requirements of each individual database. Regardless, there remains a gap in normalization when it comes to these distinct attributes, as failure to limit them could cause potential anomalies as well as issues for those attempting to process data. Thus one should consider the implementation of a new NF, Limited Distinct Normal Form (LDNF), which attempts to complement existing NFs whilst rectifying issues regarding distinct attributes by limiting them.

# 2 NORMALIZATION

Normalization was first described by Codd [1970] where he outlined the process of first normal form and its purpose in database systems. Codd [1970] had observed undesirable structural properties when it came to relational databases [1], and as a result, would come to develop a series of three normal forms which today comprise the basis of the normalization process. By applying these series of rules (NFs), thereby 'normalizing' a database, one aims to reduce data redundancy and uphold data integrity.

First normal form (1NF) requires that all fields include only atomic values, which are values that can not be separated into smaller constituent fields [2]. For instance, consider the following table:

Table 1: Scientists

| SCIENTIST (PK) | DISCOVERIES |
|---|---|
| Newton | Calculus, theory of gravity, laws of motion |
| Bohr | Bohr model of the atom |

The discoveries associated with each scientist have the potential to hold multiple values (such is the case for Newton), and thus these values are non-atomic, making it difficult to query such data. To resolve this, the following could be implemented:

Table 2: Scientists in 1NF

| SCIENTIST (PK) | DISCOVERIES (PK) |
| --- | --- |
| Newton | Calculus |
| Leibniz | Calculus |
| Newton | Theory of Gravity |
| Bohr | Bohr model of the atom |

Having the rows separated, as per table 2, adheres to 1NF by preventing non atomic values. The key in the example above is composite (i.e. made up of both the SCIENTIST and DISCOVERY) which prevents non atomic values as a new record can always be added with a unique combination of the discovery and scientist (hence why both Leibniz and Newton can be attributed to calculus without having to use non atomic values).

1NF also requires that the row order of a given table has no significance because the system will likely fail to function should the ordering conditions ever be replaced [2].

Second normal form (2NF) requires that attributes have no partial dependency. In short, a database is in 2NF if it is in 1NF and if all non-prime attributes are fully dependent on the candidate key [3]. Consider that a new non-key attribute, PLACE OF BIRTH, was added to table 2. This attribute is dependent on the SCIENTIST, but not on the DISCOVERY, and thus is said to be partially dependent on the primary key. Partial dependency can result in a series of anomalies arising, notably update, insertion, and deletion anomalies [1]. To put the above table into 2NF, and avoid partial dependency, a relational table should be made in which the primary key of this new table is SCIENTIST and the PLACE OF BIRTH is added as a non-key attribute. In this table, SCIENTIST → PLACE OF BIRTH, where → denotes functional dependency (everything on the right of the arrow is functionally dependent on the attributes on the left side of the arrow). The original table should have the PLACE OF BIRTH removed, and the SCIENTIST attribute should now also act as a foreign key. Both attributes in the original table once again depend on each other and still act as a composite key. These relational tables are now in 2NF.

In addition to partial dependency, transitive dependency may also result in data anomalies. Third normal form (3NF) requires that the database is in 2NF and that no non-prime attributes are transitively dependent on any candidate key [3]. Consider the following table:

Table 3: Scientists and Birthplace

| SCIENTIST (PK) | COUNTRY OF BIRTH | CITY OF BIRTH |
| --- | --- | --- |
| Newton | England | Woolsthorpe-by-Colsterworth |
| Bohr | Denmark | Copenhagen |

In this table all attributes are fully dependent on the primary key (SCIENTIST), fulfilling 2NF, however CITY OF BIRTH is transitively dependent on COUNTRY OF BIRTH. If Newton's country of birth were to be updated, then the city may also need to be updated to reflect this change. Such failure in updating between both attributes could lead to data inconsistencies. To resolve this, a new table could be created which contains both COUNTRY OF BIRTH and CITY OF BIRTH - in which COUNTRY OF BIRTH → CITY OF BIRTH. The COUNTRY OF BIRTH would now act as a primary key in the new table and a foreign key in the original table (CITY OF BIRTH would be removed from the original table). This means that no non-prime attribute is dependent on any other non-prime attribute and the data is now in 3NF.

Further normalization forms also exist. Boyce-Codd Normal Form (BCNF) is an adaptation of 3NF, and is more rigorous than its predecessor. It requires that "for every attribute collection C of R, if any attribute not in C is functionally dependent on C, then all attributes in R are functionally dependent on C" [4]. In practice, it means that no attributes are transitively dependent on any candidate key. This is different to 3NF in which only non-prime attributes are not to be transitively dependent on the candidate key. Nested list normal form (NLNF) builds upon this form and attempts to further account for anomalies that may arise due to functional dependencies in nested attributes [12].

Fourth normal form (4NF) requires that a database is in BCNF and all multivalued dependencies in a table must only be multivalued on the primary key [8]. Subsequently, fifth normal form (5NF) requires that the data be in 4NF and that tables must not be able to be described by the joining of other tables (i.e. tables must not be able to be reconstructed with information present in other tables) [10].

Domain-key normal form (DKNF) is an interesting NF which stipulates that a relation is in DKNF if by enforcing domain and key dependencies, every constraint of the relation will be automatically enforced [9]. It differs from other NFs by removing constraints that violate relation integrity rather than outlining a procedure to mitigate dependencies [9]. Furthermore, unlike other NFs, DKNF takes a more theoretical approach rather than a practical one.

Sixth normal form (6NF) is currently the highest form of normalization. For a relation to be in 6NF it must have no nontrivial join dependencies [6]. There also exists, elementary key normal form (EKNF) [15] and essential tuple normal form (ETNF) [5] which are positioned between the other six normal forms.

Most of these various NFs focus largely on the same issue; functional dependencies [4, 8, 12, 15]. As an extension of this some forms focus on multivalued dependencies [8] or join dependencies [10] which ultimately stem from the concepts of functional dependencies [1]. However, this normalization approach fails to consider the fundamental properties of attributes which if not normalized may also lead to inconsistencies. Thus, it is important to consider not only an approach to normalization from the relationship level, but also at the attribute level.

## 3 LIMITED AND DISTINCT ATTRIBUTES

On the subject of databases, a domain constraint refers to the set of valid values that any field in a given attribute may have [9, 13]. This is synonymous with domain dependencies, the term initially used by Fagin [1981] when outlining DKNF. Practically speaking, a domain constraint is present when the data type of a given attribute in a database has been set and the valid ranges are defined. An attribute labeled 'First Name' will likely be bound to be a string of characters no longer than 64 letters in length and an attribute designated 'Date Posted' would logically be constrained to the MM/DD/YY format.

In various programming languages, enumerated types (enums) are a data type which confines variables to a fixed set of predetermined values. In Structured Query Language (SQL), it is defined as a "string object with a value chosen from a list of permitted values that are enumerated explicitly in the column specification at table creation time" [14]. Generally speaking, enums may be employed in programs in order to reduce errors. This is because they restrict the possible values of a variable and make values relatively easier for users to recognise than if they were integers instead. Enums also take up less storage relative to string or text data types. For instance, in MySQL 8.0, an open source database management system, enums require 1 or 2 bytes based on the number of possible values whereas text requires 2 bytes plus the actual length of string value in bytes [14].

When observing enumerations in effect, one could consider a weather station choice in creating an enumeration for the different classifications of weather.

Table 4: Weather

| WEATHER (PK) |
| --- |
| Sunny |
| Partly Cloudy |
| Cloudy |
| Rainy |
| Snowing |
| Thunderstorm |

By forcing the weather to be classified into these distinct values, it ensures that different names for the same weather will not be inputted (e.g. prevents 'Thunderstorm' and 'Lightning Storm' from being used interchangeably). This becomes important when data analysts attempt to process data and produce analytical results, since there are now fewer categories to sift through. For instance, if a report was to be created with the number of times a classification of weather had been experienced in a year, analysts can now simply add up the limited and distinct values of the weather attribute. This prevents cases where in order to find all the classifications of 'Thunderstorm', analysts would need to sum all instances of fields which contain the value 'Thunderstorm', 'Lightning Storm', 'Electrical Storm', 'Storm', and potential misspellings of the aforementioned.

Attributes such as these will hereby be referred to as distinct attributes and will be defined simply as attributes that hold data for which can be outlined by a small, finite set of values. Furthermore, attributes which are distinct but not enforced through a limited attribute method will be referred to as non-limited distinct attributes (NLDAs).

In addition to query difficulty, NLDAs open the database up to anomaly risks. For instance, take the following table (which is in 3NF):

Table 5: Weather History

| EVENT ID (PK) | WEATHER TYPE | EVENT DATE | HAZARD SCALE |
| --- | --- | --- | --- |
| 1 | Thunderstorm | 01/01/2023 | 8 |
| 2 | Showers | 02/01/2023 | 1 |
| 3 | Lighting Storm | 03/01/2023 | 7 |
| 4 | Light Rain | 04/01/2023 | 1 |

Note. EVENT ID → HAZARD SCALE NOT WEATHER TYPE → HAZARD SCALE

Consider that all the thunderstorm weather events were to be changed to have a HAZARD SCALE of 9. Since WEATHER TYPE is a NLDA, despite thunderstorm and lightning storm being referring to the same weather conditions, the update will only change those events which have a WEATHER TYPE of thunderstorm. Thus, the third record would not be updated to have a HAZARD SCALE of 9 and would remain at 7, creating data inconsistencies. This is an update anomaly caused by the table having NLDAs.

## 4  A NEW NORMAL FORM

Currently all the existing normalization forms (as previously outlined) allow NLDAs to exist without any attempt in limiting them. Even 6NF - which is considered to be the highest normalization form - allows for NLDAs to exist within the table primarily due to the focus of dependencies. Although DKNF does in part consider domains by requiring domain constraints to be enforced, it does not require domain dependencies to be formed in the presence of NLDAs [9]. Thus a relation in DKNF can still have NLDAs present because by definition DKNF requires that a relation only have domain and key constraints, thereby neglecting attributes with domains that are not defined (as in the case of NLDAs). In an attempt to better uphold data integrity by promoting greater levels of consistency, one should consider the implementation of a new normalization method, Limited Distinct Normal Form (LDNF), which approaches normalization at a more fundamental level than traditional normal forms similar to DKNF [9].

LDNF proposes that NDLAs should be made limited and can only exist as its finite set of values. In practice, this may be most easily achieved by placing NDLAs into their own table and converting the attribute in the original table into a foreign key of the primary key of the new table. This new table will only contain the distinct attributes, and thus the original table will only be able to contain these distinct values as it is limited by the new referenced table. All previous NLDAs in the relation will from henceforth become limited distinct attributes (LDAs). Here it is important to note that LDAs are by definition a method of applying a domain constraint to a given attribute. However, the two do not always co-exist, as there are various cases of applying domain constraints where LDAs are not concerned (e.g. when applying a constraint on the range of values a number attribute can take on). An alternative method to applying LDNF is to utilize enums, applying the datatype to NDLAs in order to make them LDAs. This has the same effect as creating a new table, and in some cases could also be more desirable. It is also very likely that there exists further ways in which one can apply LDNF outside of the two mentioned, but as long as NDLAs are made LDAs, LDNF is satisfied.

*THEOREM 1: Suppose a relation, R, is in LDNF. Every distinct attribute, A, in R, will have a predefined distinct set of acceptable values, V, where no two values in V will be similar and distinct.*

*PROOF: Since LDNF requires that all values of a particular distinct attribute in R are made up of the distinct values in V, all values of A must therefore contain a value from V as A must contain a value and the only acceptable values are those from V. Hence, as V is designed to exclude similar values, there will be no values of A which are similar but distinct.* □

Unavoidably, there is likely to be some level of ambiguity regarding the bounds of what one should consider a distinct attribute. Earlier, distinct attributes were defined to be attributes which contain a small, finite set of potential values. How small a finite set of values must be in order to be classified as a distinct attribute, and thereby subject to be made limited through LDNF, is ultimately subjective and depends on the end user requirements of the database. This is similar to the concept of atomicity in 1NF, since the level in which an attribute must be separated down also ultimately depends on the level of detail required by end user requirements. For instance, the date of something may be considered to be atomic, even if it can be split further into day, month, and year, because it is not necessary to split the data further for the purposes of the database. Similarly, the rounded ages of the human population of the Earth at a given time is also a finite set of values, yet it is not necessary or practical for them to be considered distinct attributes. Hence, not all finite attributes should be considered distinct, and it ultimately depends on the goals of a database when it comes to determining which attributes should be classified as such.

A variety of situations arise in which LDNF should be applied. These include the application into fields which could contain different values despite having the same input (e.g. the colors 'silver' and 'gray' may be used interchangeably). LDNF also helps protect against data entries between American and British English (e.g. "color" and "colour") and between changes in words between nations (e.g. "togs" and "trunks"). If a relation is in LDNF then querying will also likely be more efficient. As a result of attributes only being made up of distinct values, queries can be optimized to run off these distinct values mitigating ambiguity when interpreting data which in turn improves query speed (due to reduced parameters) and accuracy. The usage of LDNF will also remove the presence of data anomalies that may have otherwise been caused by NLDAs.

*THEOREM 2: A relation, R, in LDNF will not have any data anomalies due to NLDAs.*

*PROOF: As attributes, A, in R only consist of values from V (from theorem 1), when updating a value it only needs to be updated in one place thereby preventing update anomalies due to value inconsistencies. Furthermore since V is always defined there will be no instances in which an insertion cannot be made (as concepts are always represented through V) thereby preventing insertion anomalies. Finally, deletion anomalies are prevented as values are always present in V and thus if deleted from A will not be lost. Therefore all NLDA anomalies are prevented. □*

Note that putting a relational database into LDNF does not mean that it has been decomposed as far as possible - in a similar way to 4NF [8]. For instance, applying LDNF will not remove translational dependency and hence will not be in 3NF. For this reason it is essential that a relation should be at least in 3NF before applying LDNF in order to ensure the database is conventionally normalized [11]. Systematic normalization is the primary way in which this process is carried out [11] and therefore should also be applied to LDNF (i.e. the data is normalized according to 1NF, then 2NF and so on until being normalized into LDNF).

## 5 EXAMPLES

Applications of LDNF in relational databases are useful in many scenarios and practical to implement. Consider the following table:

Table 6: Scientists Favorite Color

| SCIENTIST (PK) | FAVORITE COLOR |
|---|---|
| Newton | Grey |
| Hooke | Green |
| Planck | Red |
| Faraday | Silver |
| Edison | Gray |

Currently, there are three fields in FAVORITE COLOR that have different values but essentially refer to the same region of hue, lightness, and saturation. 'Gray' and 'Grey' are the respective American English and British English spellings of the same color. 'Gray' and 'Silver', whilst technically different colors, refer to essentially almost the same thing - and in a query which attempts to return all SCIENTIST fields who have a favorite color of 'Gray', one would realistically expect 'Faraday' since the colors are so similar. Therefore, 'Gray', 'Grey', and 'Silver' are, for the purposes of this table, to be considered the same color. Since COLOR is a distinct attribute (there are a small, finite number of color classifications that are commonly used) and it has not been limited to a set of values, this table is not in LDNF. In order to normalize the table in LDNF, FAVORITE COLOR must be given its own table, one which stores the distinct values held by the fields within FAVORITE COLOR in a single attribute. As we are only storing the distinct values, 'Grey' and 'Silver' will be removed, as they refer to the same color as 'Gray'. FAVORITE COLOR will become the foreign key of the primary key of this new table. Thus, the following will be produced:

Table 7: Scientists Favorite Color LDNF

| SCIENTIST (PK) | FAVORITE COLOR |
|---|---|
| Newton | Gray |
| Hooke | Green |
| Planck | Red |
| Faraday | Gray |
| Edison | Gray |

Table 8: Colors

| COLOR (PK) |
|---|
| Red |
| Green |
| Gray |

Since the FAVORITE COLOR attribute in the original table is now a foreign key of the COLOR primary key in the newly created table, the former is now limited to the values within the latter. This means that users can no longer input different values for the same color. The table is now in LDNF because all NLDAs (in this case, purely the FAVORITE COLOR attribute), have been limited to a predefined set of values.

Placing this table in LDNF helps to reduce a number of issues when it comes to deriving analytics. For instance, if information regarding the commonality of favorite colors amongst scientists was required, analysts would need to find the number of times a value appears in the FAVORITE COLORS attribute. Without applying LDNF this process could become far more inefficient, especially for larger databases, as there will be more values to sift through and classify under the necessary values (this is evident where we discussed three possible values for 'Gray'). Having a database in LDNF eliminates this issue since the values are now limited and classified under a singular term, thereby allowing for a more efficient extraction of data analytics.

It should be noted that at times, different limited distinct attributes within different tables may be foreign keys of the same table. For instance, if a new table was added with the attributes CELEBRITY and FAVORITE COLOR, where CELEBRITY → FAVORITE COLOR, the FAVORITE COLOR attribute would be a foreign key of the same table as the FAVORITE COLOR attribute in our table with SCIENTIST → FAVORITE COLOR. There is evidently no need to create a new set of values for the COLOR attribute of our new table, and maintaining the set of distinct values in one table allows for alterations to be made with greater efficiency.

In data normalization, anomalies are known as inconsistencies created within a database after having performed various operations on it [5, 9]. One of these is the update anomaly, which occurs due to ineffective structuring of a database in a way which requires multiple actions in order to fulfill the changing of repeated values in the same or other tables [7]. Failing to enact these successive actions will leave the repeated values only partially changed, thus causing inconsistencies within the database and allowing potential issues to arise in the future. As mentioned prior (see theorem 2), the usage of LDNF assists in removing the potential for update anomalies, specifically those which may occur due to repeated data within the same table. Consider the following:

Table 9: Employee Occupation

| EMPLOYEE ID (PK) | PLACE OF OCCUPATION |
| --- | --- |
| 1 | Amazing Company |
| 2 | Brilliant Business |
| 3 | Captivating Company |
| 4 | Brilliant Business |
| 5 | Delightful Inc |

This table is in 4NF since PLACE OF OCCUPATION is directly dependent on the primary key, EMPLOYEE ID, and there are no multivalued dependencies. However, if a business were to ever change its name - for instance if 'Brilliant Business' rebranded to 'Magnificent Business' - then there may be an update anomaly with values only updating partially - consider the case where the PLACE OF OCCUPATION field corresponding to EMPLOYEE ID '2' was updated to 'Magnificent Business', but the field corresponding to EMPLOYEE ID '4' was left unchanged. This is an update anomaly since the name of the business had only been partially updated. In order to apply LDNF, the PLACE OF OCCUPATION attribute must be limited to a set of values. Doing so produces the following:

Table 10: Employee Occupation in LDNF

| EMPLOYEE ID (PK) | PLACE OF OCCUPATION (FK) |
| --- | --- |
| 1 | Amazing Company |
| 2 | Brilliant Business |
| 3 | Captivating Company |

| EMPLOYEE ID (PK) | PLACE OF OCCUPATION (FK) |
|---|---|
| 4 | Brilliant Business |
| 5 | Delightful Inc |

Table 11: Places of Occupation

| PLACE OF OCCUPATION (PK) |
|---|
| Amazing Company |
| Brilliant Business |
| Captivating Company |
| Delightful Inc |

The table is now in LDNF since all NLDAs have been made limited. In this case, having applied LDNF removes the aforementioned potential update anomaly, as the business name would have to be updated directly in the PLACE OF OCCUPATION attribute, thus applying the change for all values in the original table.

## 6 CONCLUSION

Due to the potential anomalies that arise from non-limited distinct attributes it is important to consider a method for which this can be prevented. Limited Distinct Normal Form allows tables to be methodically structured in such a way as to remove any NLDAs thereby preventing data anomalies and improving query efficiency. Current normal forms do not attempt to limit these NLDAs and therefore in order to prevent the potential anomalies they may cause it is important to apply LDNF along with traditional normalization approaches.

One way this is achieved is by the isolation of NLDAs into their own identifying tables which are then joined to the original table. Isolated tables are structured such that they only contain sets of distinct values and thereby limit the values of the original table leading to LDAs.

In order to better understand the benefits of LDNF, future research could target the way in which LDNF improves query efficiency by running simulations. As stated, LDNF should improve query efficiency, however the specific impact and cost/benefit of LDNF should be analyzed in order to ascertain how the model could be further improved. Additionally, future work could discuss more effective ways of applying LDNF to a database.